\documentclass[prc, preprint, tightenlines, groupedaddress]{revtex4}
\usepackage{amsmath}
\usepackage{graphicx}		
\usepackage{graphics}

\begin{document}
\setlength{\topmargin}{0.1in}
\newcommand{\diff}[2]{\frac{\partial #1}{\partial #2}}
\newcommand{\diffr}[1]{\diff{#1}{r}}
\newcommand{\diffth}[1]{\diff{#1}{\theta}}
\newcommand{\diffz}[1]{\diff{#1}{z}}
\newcommand{\secdiff}[2]{\frac{\partial^2 #1}{\partial #2^2}}
\newcommand{\twovec}[2]{\left(\begin{array}{c} 
#1 \\ #2 \end{array}\right)}
\newcommand{\threevec}[3]{\left(\begin{array}{c} 
#1 \\ #2 \\ #3 \end{array}\right)}
\newcommand{\twomatrix}[4]{\left(\begin{array}{cc} 
#1 & #2 \\ #3 & #4 \end{array}\right)}
\newcommand{\threematrix}[9]{\left(\begin{array}{ccc} 
#1 & #2 & #3 \\ #4 & #5 & #6 \\ #7 & #8 & #9 \end{array}\right)}
\newcommand{\twomatrixdet}[4]{\left|\begin{array}{cc} 
#1 & #2 \\ #3 & #4 \end{array}\right|}

\title{Three Bosons in One Dimension with Short-Range Interactions I: Zero Range Potentials}


\author{Nirav~P.~Mehta}
\email[]{mehtan@jilau1.colorado.edu}
\author{James~R.~Shepard}
\email[]{james.shepard@colorado.edu}
\affiliation{Department of Physics, University of Colorado, Boulder, CO 80309}


\date{\today}

\begin{abstract}
We consider the three-boson problem with $\delta$-function interactions
in one spatial dimension.  Three different approaches are used to calculate
the phase shifts, which we interpret in the context of
the effective range expansion, for the scattering of one free particle
off of a bound pair.  We first follow a procedure outlined by 
McGuire in order to obtain an analytic expression 
for the desired S-matrix element.  This result is then compared 
to a variational calculation in the adiabatic hyperspherical 
representation, and to a numerical solution to the momentum space 
Faddeev equations. We find excellent agreement with the exact
phase shifts, and comment on some of the important 
features in the scattering and bound-state sectors.  In particular, 
we find that the 1+2 scattering length is divergent, marking the 
presence of a zero-energy resonance which appears as a 
feature when the pair-wise interactions are short-range.  Finally, 
we consider the introduction of a three-body interaction, and 
comment on the cutoff dependence of the coupling.
\end{abstract}

\pacs{}

\maketitle

\section{Introduction\label{intro}}
The three-body problem with short-range interactions has been of 
considerable interest for many years in nuclear physics~\cite{danilov, 
gribov, skorniakov_ter-martirosian, efimov1, efimov2, nielsen_etal, BHvK}.
More recently, with the realization of Bose-Einstein condensates in 
dilute alkali gases, and the ability to tune the 2-body scattering 
length for such atoms near a Feschbach resonance, it is of increasing 
interest in atomic physics~\cite{hammer, bulgac_et_bedaque, BBH, 
gasaneo_macek, esry_etal}.  In light of the relatively recent 
development of effective field theory (EFT), we revisit the old 
model of three particles in one dimension interacting via 
$\delta$-function interactions.  While this model has been 
considered previously by many authors, we believe this work 
provides some unique insights with regard to both atomic and nuclear physics. 

Consider the regime where typical nucleon momenta lie well below 
the pion mass.  In this limit, it is possible to construct a 
nonrelativistic EFT in which the pionic degrees of 
freedom are integrated out.  This leaves only nucleon fields with 
contact interactions, and higher order derivative corrections.  
Regarding such an EFT in the two-nucleon sector, a great deal of 
literature has emerged over the past decade~\cite{ksw,ksw_1,
kaplan,phillips_etal}.  More recently, there has been a 
focus on the three-nucleon sector~\cite{BHvK,
griesshammer, afnan_phillips}.  For a recent review 
see~\cite{hadrons_to_nuclei}.  Further, there are now a 
family of high precision nucleon-nucleon (NN) potential 
models which reproduce NN scattering phase-shifts up to
lab energies of 350~MeV (see~\cite{av18} and references therein).  
Each of these potentials treat the long-range portion of the interaction
in the same way via one-pion exchange, but differ in the treatment of the
less understood short-range physics.  Hence,
matrix elements of such interactions are said to be model dependent.  It is 
possible, however, to decimate the high momentum degrees of 
freedom by a sequence of renormalization group (RG) 
transformations in order to arrive at a model independent 
low-energy effective interaction~\cite{sunygroup2}.
This suggests that low-momentum potential models with only 
nucleons as explicit degrees of freedom may provide a sufficient 
description of few-nucleon systems.  Further, EFT may be 
used to systematize calculations of low-energy phenomenon, in 
principle, allowing calculations of arbitrarily high accuracy.

For atomic systems, one dimensional Bose gases are of particular
interest since phase fluctuations are enhanced.  It may 
seem that one dimensional geometries require 
a radial confinement of order the Bohr radius, but this is 
in fact not the case.  All that is required is that the energy gap 
in the transverse direction be much greater than the gap in the 
longitudinal direction~\cite{lieb_etal}.  Also, one 
dimensional geometries have been observed to display higher
critical transition temperatures to BEC~\cite{ketterle_druten}, and 
substantially reduced three-body recombination rates~\cite{tolra_etal}.  
These developments underscore the importance of the three-body problem with 
short-range interactions in one dimension.

This paper is the first of a pair which investigate EFT and 
low-momentum effective interactions in one dimension.  For 
simplicity, we consider only spinless bosons.  Scattering theory 
in one dimension plays a central role in all of our calculations.  
Of particular importance is the effective range 
expansion~\cite{schwinger, bethe}, which takes a slightly 
unfamiliar form.  We refer the reader to~\cite{effops} for 
the relevant one dimensional derivation.  

We calculate the exact symeterized S-matrix element for the 
scattering of one boson off of a bound pair, and derive an
analytic expression yielding the effective range expansion
to all orders for this 1+2 process. Having 
found an exact solution, we proceed to calculate the adiabatic 
hyperspherical potential curves in a manner similar to 
reference~\cite{gibson_etal}.  We use the eigenchannel 
R-matrix method~\cite{greene, mcrs} in order to determine 
the scattering phase-shifts, and find good agreement with 
the exact solution.  Our results for the phase-shifts, 
however, differ in a critical way from those presented 
in reference~\cite{amaya-tapia}.  We trace this disparity 
to varying definitions for the S-matrix element itself.  
We argue that our definition for the S-matrix element is 
consistent with the threshold behavior of the effective 
range expansion and with the statement of Levinson's theorem 
in one dimension~\cite{bianchi,gibson}.  Finally, we derive and 
solve (numerically) the momentum space Faddeev equations 
for the 1+2 scattering amplitude, and find excellent 
agreement with the exact result.  This approach also provides 
a convenient way to analyze the cutoff-dependence of the 
scattering amplitude and determine the running of the 
three-body coupling constant.    

\section{Exact Solution}
\label{exact}
McGuire~\cite{mcguire} has shown that when the 
masses of three identical bosons are the same, and the 
strengths of the pair-wise $\delta$-function interactions are equal, 
then the elements of the scattering matrix can be found by 
simple geometric optics.  Here, we briefly sketch his original 
arguments to calculate the S-matrix, and go one step further 
to show how this result can be examined in the context of
the effective range expansion.

We begin with the interaction expressed in hyperspherical 
coordinates $\rho$ and $\theta$ (see appendix~\ref{adhyprep}):
\begin{equation}
\label{vhypdel}
V(\rho,\theta)=\frac{c_0}{\sqrt{2}\rho}\left(\delta(|\cos{\theta}|) + 
\delta(|\cos{\theta-\pi/3}|) + \delta(|\cos{\theta+\pi/3}|)\right).
\end{equation}
In the two-dimensional plane covered by $\rho$ and $\theta$, 
this interaction is non-zero on three lines which intersect at 
angles of $\pi/3$.  Each of the resulting six regions 
corresponds to a unique ordering of the three particles along the real line.  
The elements of the scattering 
matrix are calculated by tracing a arbitrary ray through 
the potential diagram, and keeping track of the reflection 
and transmission amplitudes at each intersection.  $\mathbf{S}$ 
is then a six by six matrix which is indexed by a given 
ordering of the three particles.  The situation 
is further simplified by choosing one particular initial 
ordering and calculating the six corresponding elements 
indexed by the final ordering.  All other elements are 
readily found by permutations of the original ordering.


We write the familiar transmission and reflection amplitudes as:
\begin{align}
T=&\frac{\alpha}{\alpha+1}\\
R=&\frac{-1}{\alpha+1}
\end{align}
with
\begin{equation}
\alpha=\frac{2k\cos{\phi}}{imc_0}=ika_2\cos{\phi},
\end{equation}
where $k\cos{\phi}$ now denotes the momentum component 
of the initial ray which is normal to the surface of the 
$\delta$-function line.  The incoming ray can be traced 
through the potential diagram with the introduction of three
angles $\phi_1$, $\phi_2$ and $\phi_3$ denoting the angle 
with respect to the normal for the first, second and third 
$\delta$-function line, respectively.  
If we let $T_i=T(\alpha_i)$ 
and $R_i=R(\alpha_i)$ be indexed by the wave vector 
$k\cos{\phi_i}$, and let the initial ordering of the particles 
be (123) from left to right, then we find the elements 
of the S-matrix tabulated in Table~\ref{s-mat}.
\begin{table}[!t]
\caption{S-matrix for the $\delta$-function interaction; note there is
a common factor in each element, and we've defined: $\tilde{S}=S\cdot(\alpha_1+1)
(\alpha_2+1)(\alpha_3+1)$}
\begin{center}
\begin{tabular}{|c|c|c|}
$\langle (outgoing)|S|(incoming)\rangle$ & Amplitude & $\tilde{S}$\\
\hline
$\langle(123)|S|(123)\rangle$  &  $R_1 R_2 R_3 + T_1 R_2 T_3$ & $-1-\alpha_1\alpha_2$ \\
$\langle(213)|S|(123)\rangle$  &  $R_1 R_2 T_3 + T_1 R_2 R_3$ & $\alpha_2$ \\
$\langle(132)|S|(123)\rangle$  &  $R_1 T_2 R_3$ & $\alpha_2$ \\
$\langle(231)|S|(123)\rangle$  &  $T_1 T_2 R_3$ & $-\alpha_1\alpha_2$ \\
$\langle(312)|S|(123)\rangle$  &  $R_1 T_2 T_3$ & $-\alpha_2\alpha_3$ \\
$\langle(321)|S|(123)\rangle$  &  $T_1 T_2 T_3$ & $\alpha_1\alpha_2\alpha_3$
\end{tabular}
\end{center}
\label{s-mat}
\end{table}

Boundary conditions for fragmentation states in which two particles are
bound by $B_2 = \frac{1}{ma_2^2}$ are imposed by 
taking one component of the wave vector to be imaginary, so 
that $\alpha$ goes to $-1$.  For example, if particles $1$ 
and $2$ are a bound pair at large $\rho$, then we take 
$\alpha_1=-1$, so that $T_1$ and $R_1$  are both divergent.  
In order to evaluate the scattering amplitude, we then are free 
to set $T_1$ and $R_1$ to unity while any amplitude not containing
either $T_1$ or $R_1$ is set to zero.  In order to facilitate 
this, we label the momenta of the individual particles with 
the following kinematics:
\begin{align}
k_1=\frac{i}{a_2} + \frac{q}{\sqrt{6}} \\
k_2=\frac{-i}{a_2} + \frac{q}{\sqrt{6}} \\
k_3=-\sqrt{\frac{2}{3}}q.
\end{align}
This particular choice satisfies $k_1+k_2+k_3=0$ for 
center-of-mass coordinates, and 
$E=\frac{1}{2m}(k_1^2+k_2^2+k_3^2) = \frac{q^2}{2m}-B_2$, which defines 
$q$ in terms of the total energy.
If we consider particle 3 scattering off of a bound state of particles
1 and 2, then there are three available options.  Either there is 
total transmission and the ordering goes from [(12)3] to [3(21)] 
with direct amplitude:
\begin{equation}
\label{tno}
A^D=T_2 T_3 = \frac{\sqrt{6} qa_2 + 2i}{\sqrt{6} qa_2  - 6i},
\end{equation}
or there is rearrangement where the ordering goes from [(12)3] to either
[(23)1] or [(13)2], each of which occur with the same exchange amplitude:
\begin{equation}
\label{tyes}
A^X=T_2 R_3 = \frac{\sqrt{6}(qa_2) + 4i}{ -3i(qa_2)^2 - 4\sqrt{6}(qa_2) + 6i}.
\end{equation}
For the identical boson case, the coherent sum of these three amplitudes yields the 
desired 1+2 S-matrix element:
\begin{equation}
A^D+2 A^X=\exp{2i\delta}=1-\frac{8\sqrt{6}(qa_2)}{3i(qa_2)^2+4\sqrt{6}(qa_2)-6i}.
\end{equation}
By utilizing the relation:
\begin{equation}
\exp{(2i\delta)}-1=\frac{2i\tan{\delta}}{1-i\tan{\delta}},
\end{equation}
we obtain
\begin{equation}
\label{ert_delta_exact}
qa_2\tan{\delta}=4\sqrt{\frac{2}{3}}\frac{(qa_2)^2}{(qa_2)^2-2}.
\end{equation}
Expanding this quantity in powers of $q^2$ yields the effective range expansion
to all orders.  The crucial feature is that the first term
equal to the inverse of the scattering length is missing, indicating the presence of a zero energy
resonance.  To be more precise, we would expect $q\tan{\delta}=\frac{1}{a_3} + 
\frac{1}{2}r_3(q)^2 + \mathcal{O}((q)^4)$, but upon inspection of
Eq.~(\ref{ert_delta_exact}), we see that the three-body scattering length is infinite,
and the expansion begins with a term $\mathcal{O}((qa_2)^2)$.  
It should be stressed that this feature persists regardless of the strength of the 
$\delta$-function interaction.  There is 
a state at zero 1+2 collision energy for all attractive 
zero-range interactions, no matter the value of the scattering length.


\section{Faddeev Equation}
The Faddeev approach provides an independent way to analyze the threshold behaviour of the scattering amplitude.  While many
readers are familiar with Faddeev methods, in the interest of making the discussion self-contained, we provide a 
brief derivation of the integral equation describing 1+2 scattering in appendix~\ref{faddeev_review}.
If we define the amplitude $K(p,k; E)$ to satisfy Eq.~\ref{ieps} with the $i\epsilon$ replaced by a principal value
prescription, and include the normalization of the two-body bound state from Eq.~\ref{2bnorm}, then
we may identify $k\tan{\delta} = -2 K(k,k)$ to obtain an expression which is convenient in the context of the effective range expansion.
A manifestly three-body interaction parameterized as $2V_3/\Lambda^2$ can be included in the kernel in a straightforward manner:
\begin{equation}
\tilde{Z}(q,p,E) = \left[\frac{mE - q^2 - p^2}{(mE - q^2 - p^2) - p^2q^2} + \frac{2V_3}
{\Lambda^2}\right]\frac{4}{3a_2^2}\sqrt{-mE + \frac{3p^2}{4}}
\left[1 + a_2\sqrt{-mE + \frac{3p^2}{4}}\right].
\end{equation}
In appendix~\ref{3beft}, we present an alternative derivation of the kernel above starting from a many-body Lagrangian density.

\begin{figure}
\begin{center}
\leavevmode
\includegraphics[width=5.0in]{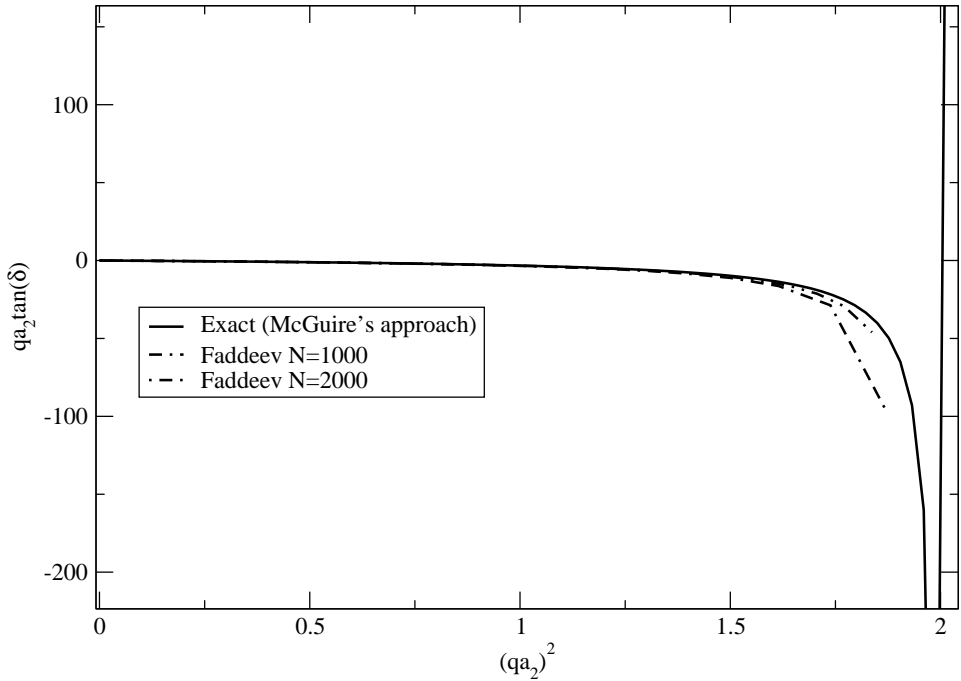}
\caption{The effective range expansion found by numerical solution to the Faddeev equation is shown.}
\label{fig:ktandel_faddeev}
\end{center}
\end{figure}

The numerical solution to principal value integral equations of the form:
\begin{equation}
\label{ls_kmat}
K(p,k;E)=U(p,k) + \mathcal{P}\int{\frac{dq}{2\pi}U(p,q)\frac{1}{k^2-q^2}K(q,k;E)}.
\end{equation}
is accomplished by letting $q=q_n$, so that the integral 
may be written in terms of matrix multiplication, and the 
principal value prescription is enforced by restricting the sum:
\begin{equation}
K_{n,m}= U_{n,m} + \sum_{l\ne n}
{dq \hspace{.1cm} U_{m,l} \frac{1}{p_n^2 - p_l^2} K_{n,l}}
\end{equation}
Which can be written more succinctly as:
\begin{equation}
\mathbf{M}\mathbf{K}=\mathbf{U}
\end{equation}
where,
\begin{align}
\label{2b_kernel}
M^n_{m,l} &= \delta_{m,l} - 
\frac{dq}{\pi} U_{m,l}\frac{1}{p_n^2-p_l^2} \\
K_{n,l} &= K(p_n,p_l).
\end{align}
Inversion of the kernel $\mathbf{M}$ is required for each 
energy (indexed above by $n$) for calculation of the on-shell K-matrix, 
and hence the phase-shifts.  When the interaction
has many high momentum components, a linear spacing of grid points 
becomes inefficient, and the inversion of the kernel 
becomes computationally cumbersome.  By performing a change 
of variable $p=\exp{(t)}$, it is possible to space the 
grid points logarithmically, facilitating the solution for such
interactions. If $p=\exp{(t)}$, then $dp=pdt$, and the 
kernel will carry an extra factor of the internal momentum $q$:
\begin{equation}
M^n_{m,l} = \delta_{m,l} - \frac{dt}{\pi} U_{m,l}\frac{p_l}{p_n^2-p_l^2} 
\end{equation}
The minimum and maximum momenta may now be chosen to define 
a domain $p \in [\exp{(t_{min})},\exp{(t_{max})}]$.  The results using the above procedure
are plotted in Fig.~\ref{fig:ktandel_faddeev}.  Clearly, as the size of the matrix is increased, the
amplitude approaches the exact result of Section~\ref{exact}.

\section{Adiabatic Curves and the Eigenchannel R-Matrix Solution}
\label{eigchan}
The final apprach involves the adiabatic representation and the eigenchannel
R-matrix method.  We refer the interested reader to appendix~\ref{adhyprep} for a 
review of these tools.
For $\delta$-function interactions, the eigenstates of the 
adiabatic Hamiltonian inside one of the six regions 
are simply solutions to the free Schr\"odinger equation.  
As long as we restrict our analysis to one of the six regions, 
the solution may be found by separation of variables and
must  be of the form~\cite{gibson_etal}
\begin{align}
\phi_0(\rho,\theta)=A_0(\rho)\cosh{(q_0\theta)} \\ 
\phi_n(\rho,\theta)=A_n(\rho)\cos{(q_n\theta)} 
\end{align}
We must treat $\phi_0$ as a special case since it represents
the only channel with a two-body bound state.
The  eigenvalue is found by demanding continuity 
of the wavefunction at the $\delta$-function surface.  If we
restrict our analysis to the region $\theta \in (0,\pi/3)$, then integration
of the Schr\"odinger equation from $\pi/6-\epsilon$ to $\pi/6+\epsilon$
leads to the following transcendental equations for $q_n$
\begin{align}
q_0\tanh{(\frac{q_0\pi}{6})} = \frac{\sqrt{2}\rho}{a_2} \\
q_n\tan{(\frac{q_n\pi}{6})} = -\frac{\sqrt{2}\rho}{a_2} 
\end{align}
Normalization of $\phi_n$ gives:
\begin{align}
A_0(\rho)=\left[\pi + \frac{3}{q_0}\sinh{\left(\frac{q_0\rho\pi}{3}\right)}\right]^{-\frac{1}{2}} \\
A_n(\rho)=\left[\pi + \frac{3}{q_n}\sin{\left(\frac{q_0\rho\pi}{3}\right)}\right]^{-\frac{1}{2}} 
\end{align}
The adiabatic potential $U_n$ is related to $q_n$ by:
\begin{align}
U_0(\rho)=\frac{-q_0^2}{2m\rho^2} \\
U_n(\rho)=\frac{q_n^2}{2m\rho^2}
\end{align}

\begin{figure}
\begin{center}
\leavevmode
\includegraphics[width=5.0in]{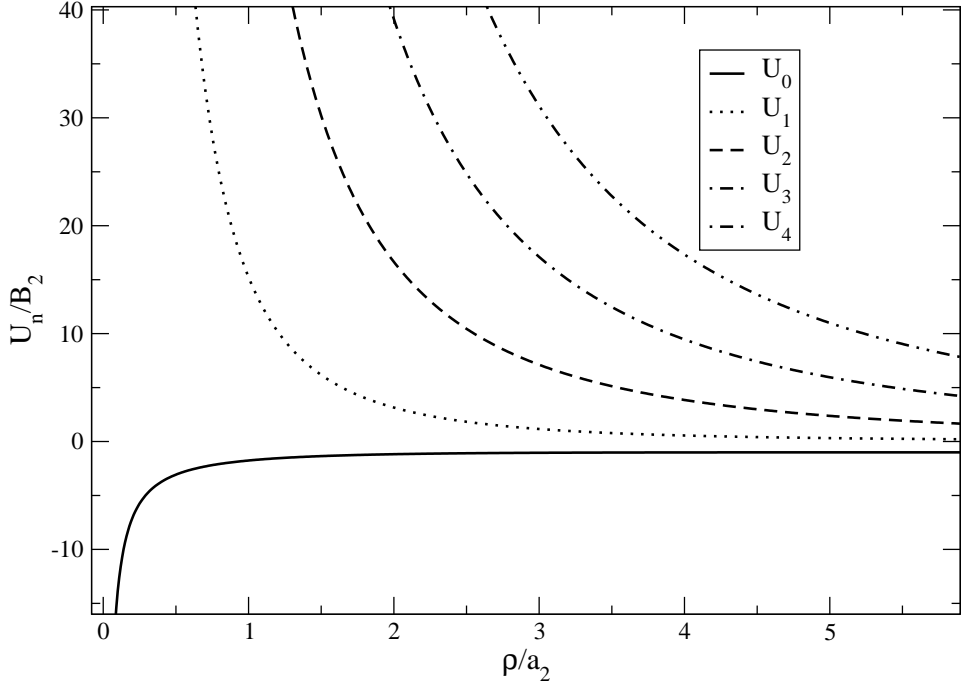}
\caption{The adiabatic potential curves for the $\delta$-function interaction are shown.}
\label{fig:delta_adcurves}
\end{center}
\end{figure}

We've verified that our potential curves are in agreement with those presented
in~\cite{gibson_etal}.  We provide a plot of the first few in 
Fig.~\ref{fig:delta_adcurves}.  As expected, there is only one attractive
channel which is open below the dimer (two-body bound state) breakup threshold.
This model supports one true bound state with energy $E=-4B_2$.  We've found that 
a calculation with one adiabatic channel under-binds this state by about $0.03B_2$.
Further, the inclusion of more coupled adiabatic channels does not serve to improve
our result.  A calculation without the diagonal coupling term gives a second
bound state at $-1.00213B_2$, very close to threshold.  When the repulsive diagonal coupling
is included, this bound state is no longer supported and its eigenenergy is above
the two-body binding.  The presence of such a state is of course consistent with the fact that the exact scattering
solution indicates a divergent scattering length.  A table of these results is shown in Table~\ref{table:delta_binding}.
\begin{table}[!t]
\caption{Eigenenergies for the $\delta$-function model 
using the adiabatic representation.  All energies are in units of the two-body binding $B_2$.}
\begin{center}
\begin{tabular}{|c|c|c|c|c|c|}
Level & Exact & 1 channel no $Q_{00}$ & 1 channel with $Q_{00}$ & 5 channels full calculation & Faddeev\\
\hline
1 &  -4         & -3.96902 & -3.96106 & -3.96106 & -3.9998\\
2 &  -1         & -1.00213 & -0.99587 & -0.99587 & -1.0000
\end{tabular}
\end{center}
\label{table:delta_binding}
\end{table}

\begin{figure}
\begin{center}
\leavevmode
\includegraphics[width=5.0in]{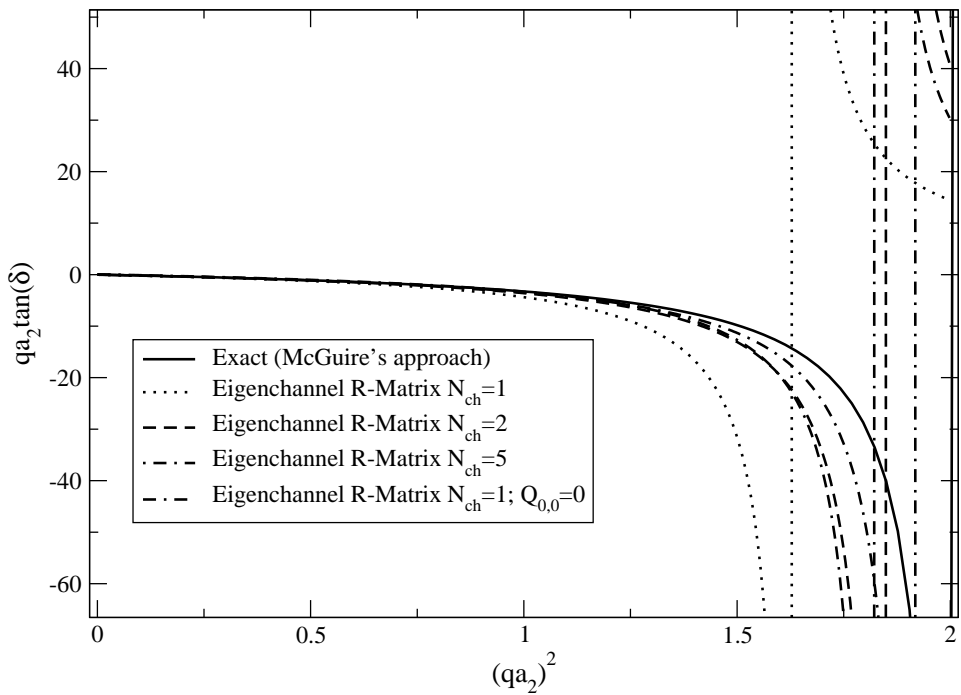}
\caption{The effective range expansion for 1+2 scattering with $\delta$=function interactions
calculated using the Eigenchannel R-Matrix method is shown.  A calculation with five coupled channels is well converged such that 
the inclusion of more channels does alter the results.}
\label{fig:ktandel_eigchan}
\end{center}
\end{figure}

For the case in question, the matrix elements in Eq.~\ref{lambdamat} and Eq.~\ref{gammamat} 
take the form (compare with Eq.~(\ref{ad_rep_SE})):
\begin{equation}
\Lambda_{m,n}=\rho_0 F_m(\rho_0)F_n(\rho_0) 
\end{equation}
\begin{align}
\Gamma_{m,n}&=\int_0^{\rho_0}
{\rho d\rho  \hspace{0.05in}\delta_{m,n} 
\left[
-\diff{F_m}{\rho} \diff{F_n}{\rho}
	+ F_m F_n (k^2 - 2mU_n(\rho)) 
\right]} \notag \\
&+ \int_0^{\rho_0}{
\rho d\rho \left[
 P_{m,n}(\rho) F_m \left(2\diff{}{\rho} + \frac{1}{\rho} \right) 
F_n + Q_{m,n} F_m F_n \right]}
\end{align}
We choose a set of $7^{\text{th}}$ order b-splines as our basis set in the expansion for the
the functions $F_n(\rho)$.  Basis splines have proven to be a versatile and efficient
basis set for a wide variety systems~\cite{vanderhart, johnson_etal, bortolotti_bohn}.  See~\cite{deboor} 
for mathematical details, fast algorithms and fortran code.  The results presented 
in Fig.~\ref{fig:ktandel_eigchan} and Table~\ref{table:delta_binding} are for a 
set of $40$ splines with a quadratic distribution of knot points over the region $\rho \in [0,20\:a_2]$.

In the scattering sector, there are a number of subtleties involved with the matching 
of the wavefunction in the asymptotic region.  If the diagonal coupling term
is ignored, then the Schr\"odinger equation in the ground state channel for $\rho\rightarrow \infty$ 
takes the form:
\begin{equation}
\left[-\secdiff{}{\rho} - \frac{1}{\rho}\secdiff{}{\rho} + k_0^2 -k^2\right]F_0(\rho) = 0
\end{equation}
This is recognized as the zeroth order Bessel's equation, and hence $f$ must be of the form:
\begin{equation}
F_0(\rho)=AJ_0(q\rho) + BY_0(q\rho)
\end{equation}
where $q^2 =  k^2-k_0^2$.  If the diagonal coupling is included, then the solution must satisfy
\begin{equation}
\left[-\secdiff{}{\rho} - \frac{1}{\rho}\secdiff{}{\rho} + k_0^2 -k^2 - Q_{00}(\rho)\right]F_0(\rho) = 0.
\end{equation}
$Q_{00}$ behaves asymptotically as $1/4\rho^2$, meaning that the solutions must now be fractional order
Bessel functions,
\begin{equation}
F_0(\rho)=AJ_{1/2}(q\rho) + BY_{1/2}(q\rho).
\end{equation}

Reference~\cite{amaya-tapia} defines the scattering matrix in terms of the 
outgoing wavefunction as
\begin{equation}
F_0(\rho)\rightarrow \sqrt{(q\rho)}(H^{(2)}_{1/2}(q\rho) + e^{2i\delta'}H^{(1)}_{1/2}(q\rho)),
\end{equation}
which, ignoring overall normalization, and taking $q\rho\gg 1$, can be written
\begin{equation}.
F_0(\rho) \rightarrow e^{-iq\rho} - e^{2i\delta'}e^{iq\rho}
\end{equation}
Reference~\cite{amaya-tapia} asserts that the minus sign appearing above adds an extra $\pi/2$ to the
phase-shift, so that the total phase shift starts at $3\pi/2$.  This assertion would
indeed be consistent with Levinson's theorem in three-dimensions, where one would
obtain a $\pi$ from the known bound state and a $\pi/2$ from the zero-energy resonance.
However, in one spatial dimension, we note that there is no additional $\pi/2$ for the
zero-energy resonance.  The statement of Levinson's theorem in 
one dimension for the even parity solution takes the form~\cite{bianchi,gibson}
\begin{align}
\lim_{k\rightarrow 0}\delta_e &= (n_e - 1/2)\pi & & \text{noncritical case} \\
\lim_{k\rightarrow 0}\delta_e &= n_e\pi & & \text{critical case}.
\end{align}
The critical case applies when there is a zero-energy resonance.
The statement for the odd-parity solution is identical to that for three-dimensional
S-waves:
\begin{align}
\lim_{k\rightarrow 0}\delta_o &= n_o\pi & & \text{noncritical case} \\
\lim_{k \rightarrow 0}\delta_o &= (n_o + 1/2)\pi & & \text{critical case}
\end{align}
We are concerned only with the even parity case.  
We define our scattering matrix in the following fashion.  We require $S$ to 
be the coefficient multiplying the outgoing $e^{iq\rho}$ in the limit $q\rho \gg 1$.
In terms of the out-going wavefunction, this means
\begin{equation}
\label{outgoing_f}
F_0(\rho) \rightarrow \frac{1}{\sqrt{q\rho}}(e^{-iq\rho} + e^{2i\delta}e^{iq\rho}).
\end{equation}
We justify our choice by considering the limit $\xi_1 \ll \xi_2$, which is appropriate when particles 1 and 2 are
bound and particle 3 is far away,  in which case, $\rho \approx \xi_2(1+\frac{1}{2}\frac{\xi1^2}{\xi_2^2} ...)$.
In this way, the product $F_0(\rho)\phi_0(\rho,\theta)$ represents a 2-particle bound
state in one relative coordinate and an oscillatory wave in the second relative coordinate, 
symeterized over all permutations of particles.  Note also that the extra factor of $1/\sqrt{q\rho}$ appearing
in our asymptotic solution is different from the convention of reference~\cite{amaya-tapia}; this is because we 
choose to work with the full wavefunction instead of the reduced wavefunction, hence our integration measure 
remains $\int{\rho \: d\rho \: d\theta}$.

In terms of standing wave solutions, the definition of the phase-shift above is consistent with
the following expression involving $\tan{(\delta)}$:
\begin{equation}
\label{real_outgoing_f}
F_0(\rho)\rightarrow \frac{1}{\sqrt{q\rho}}(\cos{(q\rho)}-\tan{(\delta)}\sin{q\rho})
\end{equation}
The form in Eq.~(\ref{real_outgoing_f}) leads to the desired product state corresponding to 1+2 scattering in one dimension.
By identifying 
\begin{equation}
\label{match1}
b=-\diff{\ln{F_0(\rho)}}{\rho},
\end{equation}
one may easily solve for $\tan{\delta}$.
An alternative argument may be formulated by simply noting that the 
normalization condition $\int{\frac{d\theta}{2\pi}\phi_0^2(\rho,\theta)}=1$ requires that $\phi_0$ scale like $\sqrt{\rho}$.
This means that the full wavefunction is proportional to the quantity $\sqrt{\rho}F_0(\rho)$.  If we consider the wavefunction at
a particular angle, and let $\Psi(\rho,\theta=\theta_0) \rightarrow A\cos{(q\rho + \delta)}$, which is
the proper asymptotic form for an even solution, then
\begin{equation}
\label{match2}
\diff{\ln{\Psi(\rho,\theta=\theta_0)}}{\rho}=-b + \frac{1}{2\rho} = -q\tan{(q\rho+\delta)}.
\end{equation}
This expression is entirely equivalent to the matching condition Eq.~\ref{match1}.
We note that our phase shift $\delta$ is related to $\delta'$ by
\begin{equation}
\tan{\delta}=\frac{-1}{\tan{\delta'}}
\end{equation}
Clearly, this will alter the behavior at threshold.  We note that with our definition,
the presence of a zero energy resonance is consistent with the threshold behavior 
of the effective range expansion, namely that the 1+2 scattering length is divergent.

\section{Discussion}
\begin{figure}
\begin{center}
\leavevmode
\includegraphics[width=5.0in]{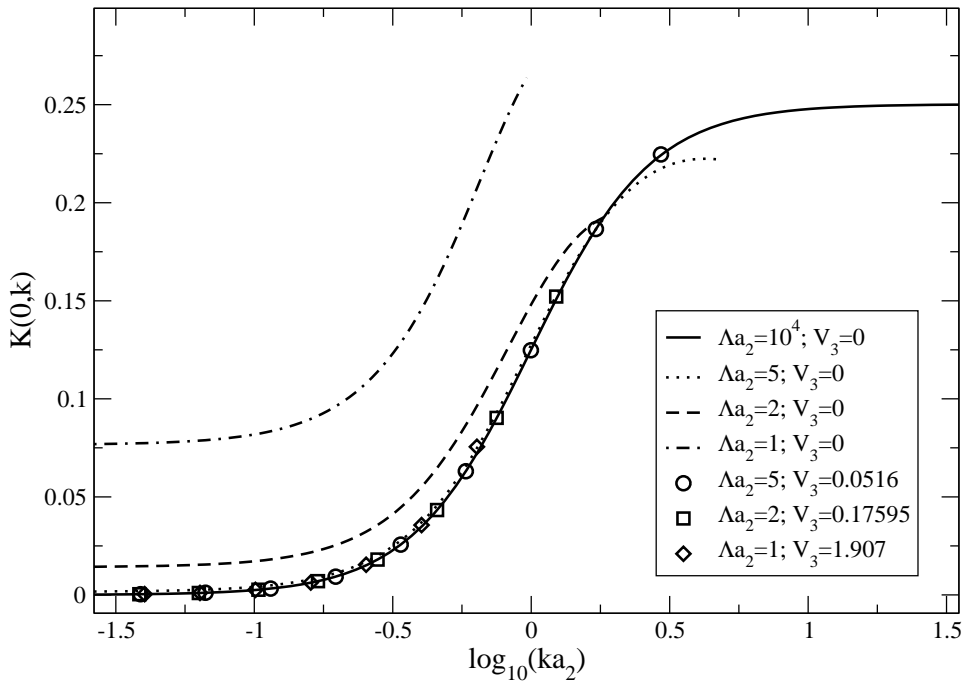}
\caption{The off-shell amplitude for our one dimensional model is shown.  The black curve which should be
considered the ``exact'' result is for $\Lambda a_2=10^4$.  As the cutoff is lowered to $\Lambda a_2=5$, a 
three-body term with $V_3=-0.0516$ brings $K(0,0)$ into agreement with the exact result.  As the cutoff
is lowered further to $\Lambda a_2=2$, a three-body term with $V_3=-0.17595$ is required.  Finally, $V_3=-1.907$ produces the 
exact $K(0,0)$ for $\Lambda a_2=1$.}
\label{fig:1DBHvK5}
\end{center}
\end{figure}

The most striking feature of these results is the presence of the zero-energy 
resonance marked by the divergent scattering length.  As the cutoff is lowered and
the range of the interaction becomes finite, the scattering amplitude in the
limit $k\rightarrow 0$ becomes nonzero.  This behavior is illustrated in the 
off-shell amplitude calculated numerically via Eq.~\ref{ieps} (except with a principal value prescription) 
shown in Fig.~\ref{fig:1DBHvK5}, which is the 1D analog of Fig.~5 appearing in reference~\cite{BHvK}.  
There is clearly a fixed point in the
limit $\Lambda \rightarrow \infty$.  The presence of the zero-energy resonance is further substantiated by
the variational calculations of Section~\ref{eigchan}.  When the repulsive second-derivative 
coupling $Q_{0,0}(\rho)$ is omitted, the interaction supports a second bound state with
energy $E=-1.00213 B_2$.  When $Q_{0,0}(\rho)$ is included, the bound state is no longer supported and a
solution to Eq.~\ref{stand_var} gives $E>-B_2$, consistent with the upper bound theorem.  These results are also
consistent with the threshold behaviour of $qa_2\tan{\delta}$ shown in Fig.~\ref{fig:ktandel_thresh}.  
The Eigenchannel R-Matrix method gives $\lim_{qa_2 \rightarrow 0}{qa_2\tan{\delta}} \approx 0.065$ when the diagonal coupling
is excluded, and $\lim_{qa_2 \rightarrow 0}{qa_2\tan{\delta}} \approx -0.008$ when it is included.
Finally, the exact effective range expansion calculated in Section~\ref{exact} indicates a divergent scattering length in perfect 
agreement with the numerical calculations.  It is again of considerable note that the zero energy resonance is present regardless of the
value of the $\delta$-function coupling, or equivalently the two-body scattering length.  It appears at exactly zero relative energy
in the 1+2 system as long as the two-body interactions are of zero range, and moves away from threshold as the interactions become of
finite range.  

\begin{figure}
\begin{center}
\leavevmode
\includegraphics[width=5.0in]{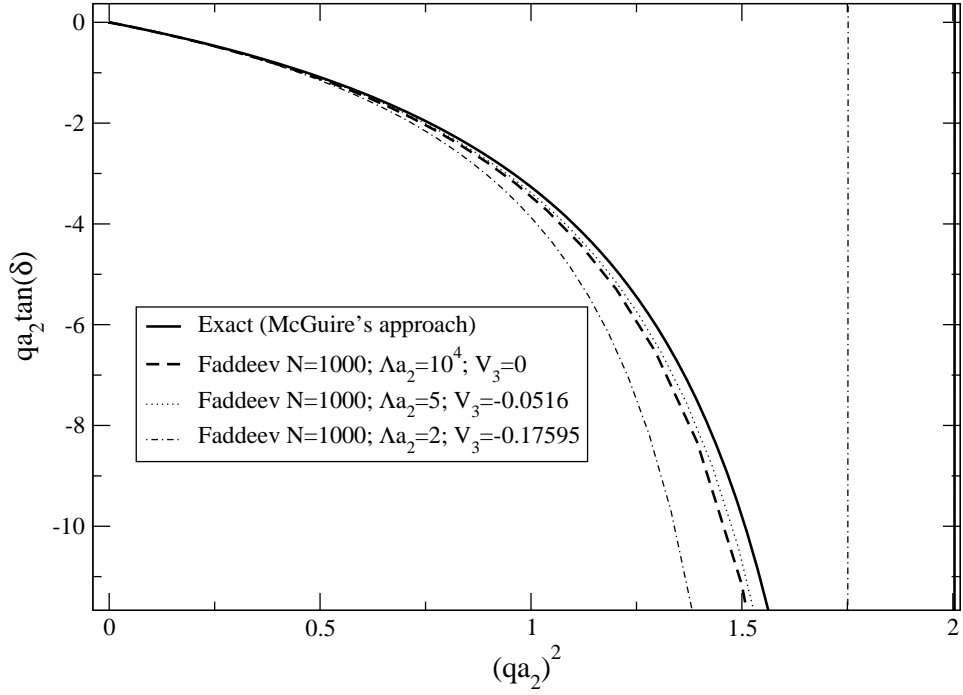}
\caption{The on-shell amplitude for our one dimensional model is shown for various values of the cutoff.  As the
cutoff is lowered, a three-body contact interaction of natural size is introduced to fix the 1+2 scattering length.}.  
\label{fig:ktandel_EFT}
\end{center}
\end{figure}

\begin{figure}
\begin{center}
\leavevmode
\includegraphics[width=5.0in]{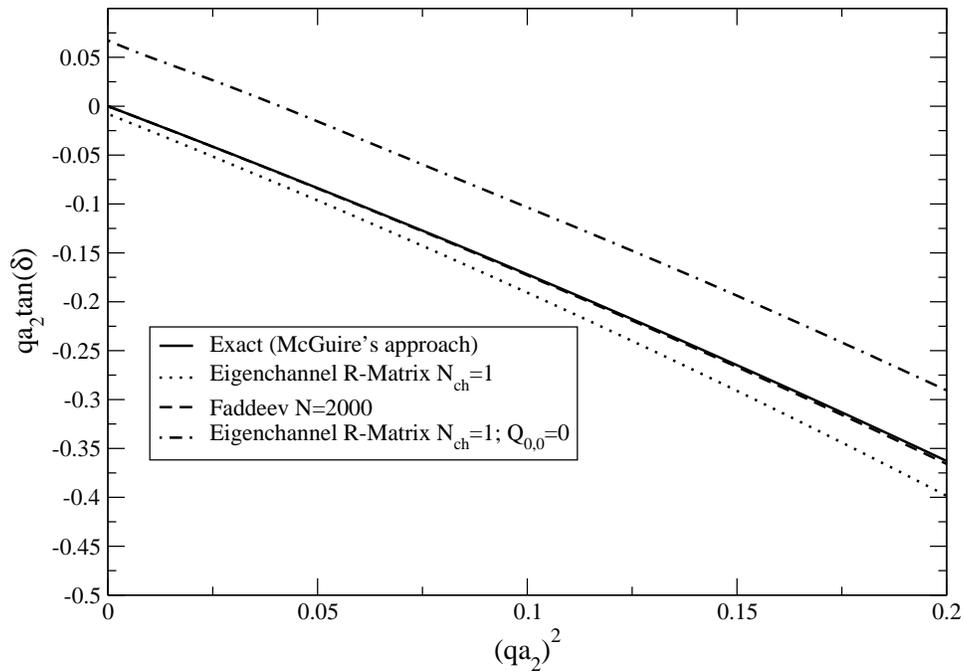}
\caption{The effective range expansion near the region $qa_2 \rightarrow 0$ is shown for various calculations.}
\label{fig:ktandel_thresh}
\end{center}
\end{figure}

It is now instructive to consider the on-shell results when various cutoffs are enforced in Eq.~\ref{ieps}.
As the cutoff is lowered, we require that $1/a_3=0$ ($a_3$ being the 1+2 scattering length).  This quantity is 
proportional to the scattering amplitude satisfied by the principal value version of Eq.~\ref{ieps}.  With
the introduction of a three-body interaction, the cutoff dependence of the amplitude can be absorbed into 
the three-body coupling $V_3$, yielding a largely cutoff invariant amplitude as seen if Fig.~\ref{fig:1DBHvK5} 
and Fig.~\ref{fig:ktandel_EFT}.

This paper has treated three bosons only at zeroth order in EFT.  In a second paper, we shall extend the
analysis to include the two-body effective range and shape parameter.  Predictions from the resulting EFT will be compared to
calculations using a realistic phenomenological NN interaction.  Three-nucleon observables will 
serve as a testing ground for the effective theory.

\begin{acknowledgments}
NPM would like to thank Chris Greene for many helpful discussions regarding the eigenchannel
R-matrix method and the adiabatic hyperspherical representation, Daniele Bortolotti for discussions on
b-splines, and the INT in Seattle for their hospitality.  NPM and JRS would like to thank Jae Park for
discussions on 1D Bose gases.
\end{acknowledgments}

\appendix
\section{Review of the Adiabatic Hyperspherical Representation and the Eigenchannel R-Matrix method}
\label{adhyprep}
The essential strategy for solving the Schr\"odinger equation in coordinate space
is to transform the partial differential equation (PDE) into a set of coupled ordinary 
differential equations (ODEs).  While there are a variety of representations that realize
this goal, the one best suited to the present problem is the adiabatic 
hyperspherical representation~\cite{macek} (see also~\cite{fano}).

Let us first introduce the appropriate relative and hyperspherical coordinates.  
Jacobi coordinates in one dimension for equal masses are defined via
\begin{equation}
\threevec{X}{\xi_1}{\xi_2}=\threematrix{\frac{1}{\sqrt{3}}}{\frac{1}{\sqrt{3}}}{\frac{1}{\sqrt{3}}}
				{\frac{1}{\sqrt{2}}}{\frac{-1}{\sqrt{2}}}{0}
				{\frac{1}{\sqrt{6}}}{\frac{1}{\sqrt{6}}}{\frac{-2}{\sqrt{6}}}
\threevec{x_1}{x_2}{x_3}
\end{equation}
where $X$ marks the position of the total center-of-mass, $\xi_1$ is 
the relative coordinate for the first two particles , and $\xi_2$ is 
the relative coordinate between third particle to the center-of-mass of 
the first two.  Hyperspherical coordinates in one dimension are simply
circular polar coordinates:
\begin{align}
\xi_1&=\rho\cos{\theta}\\
\xi_2&=\rho\sin{\theta}\\
\rho^2&=\xi_1^2 + \xi_2^2
\end{align}
Here, $\rho$ is a measure of the general size of the system.  At 
small values of $\rho$, all three particles are in close proximity,
while at large values of $\rho$, the situation depends on 
the angle $\theta$.  There are some values of $\theta$ that correspond 
to two of the three particles being near each other, and other values 
of $\theta$ where all three particles are far apart.  

With this transformation, the Schr\"odinger equation becomes (compare with Eq.~\ref{vhypdel})
\begin{equation}
\left\{\frac{1}{\rho}\diff{}{\rho}\left(\rho \diff{}{\rho}\right) + \frac{1}{\rho^2}
\secdiff{}{\theta} + k^2 - U(\rho, \theta)\right\}\Psi(\rho, \theta) = 0
\label{full_SE}
\end{equation}
where $k^2=2mE$ and 
\begin{align}
\label{vhyp}
U(\rho,\theta)=2m \Big( & V(\sqrt{2}\rho|\cos(\theta)|) + \notag \\ 
		& V(\sqrt{2}\rho|\cos(\theta+\pi/3)|) + V(\sqrt{2}\rho|\cos(\theta-\pi/3)|)\Big). 
\end{align}
This potential has a very high degree of symmetry:
\begin{align}
U(\rho, \theta)&=U(\rho,\theta \pm \frac{\pi}{3}) \hspace{1 in} &\text{exchange}\\
U(\rho, \theta)&=U(\rho,-\theta)			   	      &\text{parity}
\end{align}
The combination of exchange and parity result in a six-fold symmetry 
allowing the angular part of the wavefunction to be represented as a 
sum over terms proportional to $\cos{6n\theta}$.

The reduction of the PDE Eq.~\ref{full_SE} into a set of coupled ODEs is 
accomplished by expanding the wave function into a sum over different adiabatic channels:
\begin{equation}
\Psi(\rho,\theta)=\sum_n\psi_n(\rho,\theta)=\sum_n{\phi_n(\rho,\theta)F_n(\rho)}.
\label{channel_expansion}
\end{equation}
where $\phi_n$ are defined as eigenstates of the adiabatic Hamiltonian:
\begin{equation}
\label{adse}
H_{ad}(\rho,\theta)\phi_n(\rho,\theta)=k_n^2(\rho)\phi_n(\rho,\theta)
\end{equation}
with
\begin{equation}
\label{Had}
H_{ad}(\rho,\theta) = \frac{-1}{\rho^2}\secdiff{}{\theta} + U(\rho,\theta).
\end{equation}
It is important to note that Eq.~\ref{adse} depends only parametrically on $\rho$.  
We impose boundary conditions such that $\phi_n$ is even at $\theta=0$ 
and $\theta=\pi/6$.  
Inserting the expansion in Eq.~(\ref{channel_expansion}) into a variational expression
of the form 
\begin{equation}
k^2=\frac{\int_V{dV\:\Psi^\dagger(-\nabla^2+U)\Psi}}{\int_V{dV\:\Psi^\dagger\Psi}}
\label{stand_var}
\end{equation}
and demanding that the solution be stationary with 
respect to variations in the functions $F_m^\dagger$, one arrives at the following
matrix equation:
\begin{equation}
(\mathbf{T}+\mathbf{U})\mathbf{F}=k^2\mathbf{F}.
\label{ad_rep_SE}
\end{equation}
We've defined
\begin{equation}
\label{ke}
T_{m,n}=-\delta_{m,n}\left(\secdiff{}{\rho}+\frac{2}{\rho}\diff{}{\rho}\right) - 
P_{m,n}(\rho)\left(2\diff{}{\rho}+\frac{2}{\rho}\right)-Q_{m,n}(\rho)
\end{equation}
and
\begin{equation}
U_{m,n}(\rho)=\delta_{m,n}(k^2_m(\rho)).
\end{equation}
Also, we've introduced the non-adiabatic channel couplings defined as
\begin{equation}
\label{first_deriv_coup}
P_{m,n}(\rho)=\Big\langle\phi_m\Big|\diff{}{\rho} \phi_n\Big\rangle,
\end{equation}
\begin{equation}
\label{sec_deriv_coup}
Q_{m,n}(\rho)=\Big \langle\phi_m\Big|\secdiff{}{\rho} \phi_n \Big\rangle.
\end{equation}
The solution of the the adiabatic equatation Eq.~\ref{adse}. 
Eq.~(\ref{Had}) and the evaluation of the first 
and second derivative couplings accounts for the vast majority 
of the computational effort in solving the three-body problem using this approach.  
The first-derivative couplings $P_{m,n}$ 
can be evaluated using a Feynman-Hellman like argument.  For a 
$\rho$-parameterized system defined by $H(\rho)\phi=E(\rho)\phi$, the 
Feynman-Hellman theorem states 
\begin{equation}
\frac{dE}{d\rho}=\Big\langle\phi\Big|\diff{H}{\rho}\Big|\phi\Big\rangle.
\end{equation} 
This relation can be used to find that
\begin{equation}
\Big\langle\phi_m\Big|\diff{}{\rho}{\phi_n}\Big\rangle=\frac{\langle\phi_m\Big|\diff{H}{\rho}\Big|\phi_n\rangle}
{k^2_n(\rho)-k^2_m(\rho)}
\end{equation}
for $m\ne n$, and vanishes for m=n.  The second-derivative 
couplings, $Q_{m,n}$, may be readily calculated by noting:
\begin{equation}
Q_{m,n} = [P^2]_{m,n} + \left[\diff{P}{\rho}\right]_{m,n}.
\end{equation}
Use of the above relation, however requires calculating the first-derivative
couplings between many channels so that the square of $P$ converges.  For
calculations involving a single adiabatic channel, it is more convenient to 
estimate $Q_{0,0}$ by using a 3-point or 5-point rule.  This involves solving
the adiabatic Hamiltonian at three consecutive nearby values of the 
parameter $\rho$, and calculating the second derivative 
numerically before evaluating the inner product.

In order to calculate wavefunctions and scattering amplitudes in the scattering sector, we use
the Eigenchannel R-matrix approach~\cite{greene,mcrs}, which
is a variational calculation for minus the log-derivative of 
the wave function on the surface $S$ of some reaction volume $V$.  
More precisely, this method finds variational solutions that have a {\it constant} 
log-derivative on the surface such that $\diff{\Psi}{n} + b\Psi=0$.

Starting with Eq.~\ref{stand_var},
we define $b=-\frac{\partial{ln(\Psi)}}{\partial{\hat{n}}}$, where 
$\hat{n}$ represents the unit normal vector to the reaction surface $S$ 
($\hat{n}=\hat{\rho}$ in our case); application of Green's theorem 
to the kinetic energy term in Eq.~(\ref{stand_var}) allows us to 
write an expression for $b$ at a fixed $k^2$.
\begin{equation}
b=\frac{\int_V{dV\left[(-\vec{\nabla \Psi^\dagger}\cdot\vec{\nabla \Psi}) + 
\Psi^\dagger(k^2-U)\Psi\right]}}{\int_S{dS\:\Psi^\dagger\Psi}}
\label{varb}
\end{equation}
Note that we were able to factor $b$ out of the surface integral 
in the denominator only because the desired solution has a constant 
log-derivative on the surface $S$.  Eq.~(\ref{varb}) is an identity 
obeyed by exact eigenstates of the Schr\"odinger equation 
Eq.~(\ref{full_SE}) that have a constant $b$ 
on $S$.  By taking the first-order variation of this expression 
with respect to small deviations in $\Psi$, this expression can 
be shown to be a variational expression for $b$.

In the adiabatic representation, we expanded the wavefunction according to Eq.~\ref{channel_expansion}.
Now we expand $F_n=\sum_\alpha{c_{n,\alpha} B_\alpha}$, and Eq.~(\ref{varb}) 
is cast into the form of a generalized eigenvalue equation:
\begin{equation}
\label{eigchan_mat}
b\mathbf{\Lambda}\mathbf{c}=\mathbf{\Gamma}\mathbf{c} 
\hspace{1in} b=-\diff{\ln{\Psi(\rho)}}{\rho}
\end{equation}
where
\begin{equation}
\label{lambdamat}
\Lambda_{m,n}=\int_S{\vec{dS}\cdot \hat{n} \; \psi_m \psi_n}
\end{equation}
\begin{equation}
\label{gammamat}
\Gamma_{m,n}=\int_V{dV[-\vec{\nabla \psi_m}\cdot 
\vec{\nabla \psi_n} + \psi_m(k^2-U)\psi_n]} 
\end{equation}

\section{Derivation of Faddeev Equations}
\label{faddeev_review}
In this section, we derive the properly symeterized integral equation satisfied 
by the scattering amplitude for one free boson off
of a bound pair.  Our derivation relies heavily on the original work of 
Faddeev~\cite{faddeev}, Lovelace~\cite{lovlace} and Amado~\cite{amado}, however we 
largely hold to the notation conventions of Watson and Nuttall~\cite{watson_nuttall}.
We shall begin by calculating the two-boy scattering amplitude for a separable interaction of the 
form:
\begin{equation}
V = c_0 |g \rangle \langle g|,
\end{equation}
with momentum space matrix elements:
\begin{equation}
\langle q' |V| q \rangle = c_0 g(q')g(q).
\end{equation}
The $T$-matrix element $\langle q'|T(E)|q \rangle$ satisfies the Lippman-Schwinger equation:
\begin{equation}
T(q', q, E) = c_0g(q')g(q) + c_0g(q')\int{\frac{dq''}{2\pi}\frac{m g(q'') T(q'',q, E)}{mE - q''^2}}
\end{equation}
The solution is found by defining the energy dependent function~\cite{watson_nuttall}
\begin{equation}
h(q, E) = \int{\frac{dq''}{2\pi}\frac{mg(q'')T(q'',q, E)}{mE - q''^2}}
\end{equation}
which satisfies the algebraic equation:
\begin{equation}
h(q, E) = \left[\int{\frac{dq''}{2\pi}\frac{mg^2(q'')}{mE - q''^2}} \right][c_0 g(q) + c_0 h(q, E)].
\end{equation}
Solving for $h(q, E)$ and substituting the result into the Lippman-Schwinger equation quickly yields the 
solution:
\begin{equation}
\label{2btmat}
T(q', q, E)=g(q')g(q)\left[\frac{1}{c_0} - \int{\frac{dq''}{2\pi}\frac{m g^2(q'')}{mE - q''^2}}\right]^{-1}
\end{equation}
Taking the limit $g(q)\rightarrow 1$ is equivalent to solving the corresponding Schr\"odinger equation with contact interactions.  This 
yields a $T$-matrix element independent of $q$ and $q'$:
\begin{equation}
\label{2bdelta_tmat}
m T(E+i\epsilon) = mc_0\left[1 + \frac{mc_0}{2\sqrt{-mE-i\epsilon}}\right]^{-1}.
\end{equation}
In the center of mass frame, the energy is written $E = k^2/m$ and the 
scattering amplitude $f$ for the even parity wave is related to the on-shell $T$-matrix by:
\begin{equation}
f = \frac{1}{2}\left(e^{2i\delta}-1\right)=\frac{i\tan\delta}{1-i\tan\delta}=
\frac{m T(k,k, E)}{2ik} = \frac{\frac{mc_0}{2ik}}{1-\frac{mc_0}{2ik}}.
\end{equation}
The cross-section is a normalized probability in one dimension and is given by $\sigma=|f|^2$.

If the coupling is negative, the interaction supports a bound state.  The $T$-matrix element will exhibit a pole at the 
binding energy $E=-B_2$.  From inspection of Eq.~(\ref{2bdelta_tmat}), it is clear that this requires $B_2=mc_0^2/4=1/ma_2^2$.
The state vector for the bound state is:
\begin{equation}
|\psi_B\rangle = \sqrt{N}G_0(-B_2)|g\rangle.
\end{equation}
where $G_0(-B_2)=(-B_2-H_0)^{-1}$ is the free particle propagator evaluated at the bound state energy.  
The normalization constant is easily evaluated by contour integration to be 
\begin{equation}
\label{2bnorm}
N^{-1}=\langle g|G_0^2(-B_2) |g\rangle = \frac{m^2a_2^3}{4}.
\end{equation}


The indices in the three-body sector follow the convenient ``odd-man-out'' notation.  When considering matrix 
elements of two-body operators in the three-body state-space, 
the two-body operator $V_\alpha$ with $\alpha=1$ shall
denote the interaction between particles $2$ and $3$.  We are only concerned with internal degrees of freedom, and will
therefore work in the total center of momentum frame $p_1 + p_2 + p_3 = 0$.  Matrix elements will be taken with respect
to state vectors of the form $|p_\alpha, q_\alpha \rangle$, where $p_\alpha$ represents the momentum of the spectator particle $\alpha$, and
$q_\alpha$ represents the relative momentum of the remaining two particles.
Let $|\psi_\alpha^+\rangle = |\psi_\alpha^{(1)}\rangle + |\psi_\alpha^{(2)}\rangle + |\psi_\alpha^{(3)}\rangle$ 
describe an eigenstate of the full Hamiltonian $H$ which corresponds to an initial state $|\chi_\alpha\rangle$
with the two particles not equal to $\alpha$ forming a bound state.  The solution is found by solving the Faddeev equations for
the components $|\psi_{\alpha}^{(\beta)}\rangle$:
\begin{equation}
\label{faddeev1}
|\psi_\alpha^{(\beta)}\rangle = G_0(E)|\alpha, p_\alpha\rangle\delta_{\alpha\beta} + 
\sum_{\gamma \ne \beta}{G_0(E)T_\beta|\psi_\alpha^{(\gamma)}\rangle}.
\end{equation}
We write the two-body T-matrix in the three-body state space as:
\begin{equation}
T_\beta(E) = \int{\frac{dp_\beta}{2\pi}|\beta, p_\beta\rangle\tau_\beta\left(E-\frac{3p_\beta^2}{4m}\right) \langle \beta, p_\beta|}
\end{equation}
where $\tau_\beta(E)$ is the dimer propagator in the $\beta$ channel:
\begin{equation}
\tau_\beta(E)=\left[\frac{1}{c_0} + \int{\frac{dq}{2\pi}\frac{mg_\beta^2(q)}{q^2-mE}}\right]^{-1}
\end{equation}
For $g_\beta(q) \rightarrow 1$, $\tau_\beta(E)$ is equal to the two-body T-matrix found in the previous section.
left multiplying Eq.~(\ref{faddeev1}) by $\langle \delta, p_\delta |$ and summing over $\beta \ne \delta$ leads to:
\begin{equation}
X_{\delta,\alpha}(p_\delta, p_\alpha) = Z_{\delta, \alpha}(p_\delta, p_\alpha) + \sum_{\beta}{\int{\frac{dp_\beta}{2\pi}
Z_{\delta, \beta}(p_\delta, p_\beta)\tau_\beta\left(E - \frac{3p_\beta^2}{4m}\right)X_{\beta,\alpha}(p_\beta, p_\alpha)}}.
\end{equation}
The amplitudes $X_{\delta,\alpha}(p_\delta, p_\alpha)$ and $Z_{\delta, \alpha}(p_\delta, p_\alpha)$ are defined as:
\begin{align}
X_{\delta,\alpha}(p_\delta, p_\alpha) &= \sum_{\gamma \ne \delta}{\langle \delta, p_\delta | \psi_\alpha^{(\gamma)}\rangle.} \\
Z_{\delta, \alpha}(p_\delta, p_\alpha) &= (1-\delta_{\delta\alpha})\langle \delta, p_\delta| G_0(E) | \alpha, p_\alpha \rangle.
\end{align}
The Born amplitude $Z_{\delta, \alpha}(p_\delta, p_\alpha)$ describes the interaction mediated by the exchange of a single particle,  
and requires calculating $q_\delta$ in terms of $p_\alpha$ with $\delta \ne \alpha$. The 
kinematics for a given case must be determined by cyclic permutation of the particles~\cite{lovlace}.  For example, 
\begin{equation}
Z_{2,1}(p_2, p_1) = \frac{g_2\left(-p_1 - \frac{1}{2}p_2\right)g_1\left(p_2 + \frac{1}{2}p_1\right)}
{E-\frac{3p_1^2}{4m} - \frac{1}{m}\left(p_2 + \frac{1}{2}p_1\right)^2}.
\end{equation}
For identical bosons the quantity of interest is the symeterized amplitude given by the 
sum of the direct and exchange pieces $X(p, k; E) = X^{(D)}(p, k; E) + 2X^{(N)}(p, k; E)$.  This amplitude satisfies
the following integral equation:
\begin{equation}
\label{faddeev}
X(p, k; E) = 2Z(p, k; E) + 2\int{\frac{dq}{2\pi}Z(p, q; E)\tau\left(E-\frac{3q^2}{4m}\right)X(q,k; E)}
\end{equation}
where the Born term for $g(q) \rightarrow 1$ is given by:
\begin{equation}
Z(p, q; E) = \frac{m}{mE - q^2 - p^2 - q\cdot p},
\end{equation}
and the total energy is $E=\frac{3k^2}{4m} - B_2$.
Next we perform an angle average over the dot product. In one dimension the angle average of an 
arbitrary function $f(p\cdot q)$ is $\bar{f} = \frac{1}{2}(f(pq) + f(-pq))$, and so the Born amplitude becomes:
\begin{equation}
Z(p,q; E) = \frac{mE - q^2 - p^2}{(mE - q^2 - p^2)^2 - p^2q^2}.
\end{equation}
The desired amplitude now satisfies the integral equation:
\begin{align}
\frac{X(p,k,E+i\epsilon)}{2m} &= Z(p, k, E) \notag \\ 
&\: + 4\int{\frac{dq}{2\pi}Z(p,q,E)
\left[-a_2 + \frac{1}{\sqrt{-mE + \frac{3q^2}{4} -i\epsilon}}\right]^{-1}\frac{X(q,k,E+i\epsilon)}{2m}}.
\end{align}
It is desirable to remove the pole in the dimer propagator and bring this equation into the form
of the Lippman-Schwinger equation; to this end we define the amplitude:
\begin{equation}
\frac{\tilde{X}(p,k,E)}{k^2-p^2} = \left[-a_2 + \frac{1}{\sqrt{-mE + \frac{3p^2}{4}}}\right]^{-1}\frac{X(p,k,E)}{2m}
\end{equation}
A bit of algebra shows that the new amplitude satisfies the equation:
\begin{equation}
\label{ieps}
\tilde{X}(p,k,E+i\epsilon) = \tilde{Z}(p,k,E+i\epsilon) - \frac{4}{\pi}\int_{0}^{\infty}
{dq\:\tilde{Z}(q,p,E+i\epsilon) \frac{\tilde{X}(q,k,E+i\epsilon)}{q^2-k^2-i\epsilon}}
\end{equation}
with
\begin{equation}
\tilde{Z}(q,p,E) = \left[\frac{mE - q^2 - p^2}{(mE - q^2 - p^2) - p^2q^2}\right]\frac{4}{3a_2^2}\sqrt{-mE + \frac{3p^2}{4}}
\left[1 + a_2\sqrt{-mE + \frac{3p^2}{4}}\right].
\end{equation}
It is computationally more convenient to deal with an amplitude which is real below the breakup threshold by writing the 
above integral equation in terms of a principal value prescription using the well known formula:
\begin{equation}
\frac{1}{\omega \pm i\epsilon} = \mathcal{P} \mp i\pi\delta(\omega)
\end{equation}

\section{Alternative derivation of Eq.~\ref{faddeev}}
\label{3beft}
Bedaque et al~\cite{BHvK} have considered the 3-D three-body problem with short range interactions
in a ground-breaking paper.  It is instructive to consider their approach in a 1-D context, and that 
is the purpose of this appendix; a complete analytic sum of the series arising in perturbation theory has 
been found by Thacker~\cite{thacker}.  

\begin{figure}[!t]
\begin{center}
\leavevmode
\includegraphics[width=5.5in]{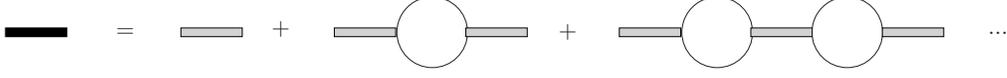}
\caption{The dressed dimer propagator determined by the geometric series of loop insertions.}
\label{fig:dimerprop}
\end{center}
\end{figure}

Consider the Feynman rules resulting from the Lagrangian density:
\begin{equation}
\label{L_nuc}
\mathcal{L} = \phi^\dagger(x) \left(i\partial_0 + \frac{\partial_x^2}
{2m} \right)\phi(x) - \frac{c_0}{2}\left(\phi^\dagger(x)\phi(x)\right)^2 - \frac{d_0}{6}\left(\phi^\dagger(x)\phi(x)\right)^3.
\end{equation}
Let us first sum the perturbative series of bubble diagrams for the two-body problem with $d_0=0$.
Let $(p_0, \vec p) = (\frac{1}{4m}(k_1-k_2)^2, k_1+k_2)$ with $|k_1|=k$
denote the two-vector in the center of momentum frame. 
The following loop integral is readily evaluated by contour integration:
\begin{align}
L&=\int{\frac{d\vec q}{2\pi}\frac{dq_0}{2\pi}\frac{i}{\frac{p_0}{2} - 
q_0 - \frac{1}{2m}(\frac{\vec p}{2}- \vec q)^2 + i\epsilon}
\frac{i}{\frac{p_0}{2} + q_0 - \frac{1}{2m}(\frac{\vec p}{2}+\vec q)^2 + i\epsilon}}\\
&=\frac{-im}{2\sqrt{-mp_0 + \frac{\vec p^2}{4} -i\epsilon}} = \frac{m}{2k},
\end{align}
and the geometric series is easily summed to reproduce the result of Section~\ref{faddeev_review}:
\begin{equation}
iA=-iT=-ic_0\left( 1 + (-ic_0L) + (-ic_0L)^2 ...\right) = 
\frac{-ic_0}{1 + ic_0L} = \frac{-ic_0}{1-\frac{mc_0}{2ik}}.
\end{equation}
Kaplan~\cite{kaplan} suggested that the Lagrangian Eq.~\ref{L_nuc} may be conveniently rewritten in terms of a dummy field 
$D$:
\begin{equation}
\label{L_D}
\mathcal{L}=\phi^\dagger \left(i\partial_0 + \frac{\partial_x^2}
{2m} \right)\phi +\Delta D^\dagger D - \frac{g}{\sqrt{2}}(D^\dagger \phi \phi + \phi^\dagger\phi^\dagger D)
+ h(D^\dagger D \phi^\dagger \phi + \phi \phi^\dagger DD^\dagger)
\end{equation}
Gaussian path integration over the 
auxiliary field $D$ shows that the couplings appearing in Eq.~(\ref{L_D}) are
related to those in Eq.~(\ref{L_nuc}) by $g^2/\Delta=c_0$ and $-3hg^2/\Delta^2=d_0$.
The bare dimer propagator is $i/\Delta$, while the sum of diagrams shown in Fig.~\ref{fig:dimerprop} yields
\begin{align}
i\Delta(p_0, \vec p)&=\frac{i}{\Delta}\left(1 + \left(\frac{-ig^2 L}{\Delta}\right) + \left(\frac{-ig^2 L}{\Delta}\right)^2 +\:...\right)\\
&=\frac{i}{\Delta+\frac{mg^2}{2}(-mp_0 + \frac{\vec{p}^2}{4}-i\epsilon)^{-1/2}}.
\end{align}

For the 1+2 integral equation, we choose the same kinematics as Bedaque et al~\cite{BHvK}.  
Let the incoming particle and dimer have two-momenta $(k^2/2m, -\vec{k})$
and $(k^2/4m-B_2, \vec{k})$, respectively.  The outgoing particle and dimer are off-shell with two-momenta
$(k^2/2m - \varepsilon, - \vec{p})$ and $(k^2/4m - B_2 + \varepsilon, \vec{p})$, respectively.
Our integral equation is identical to Eq.~(5) in~\cite{BHvK}, except that the integration measure
is $\int{ \frac{dq}{2\pi} \frac{dq_0}{2\pi} }$ for 1+1 dimensions:
\begin{figure}[!t]
\begin{center}
\leavevmode
\includegraphics[width=5.5in]{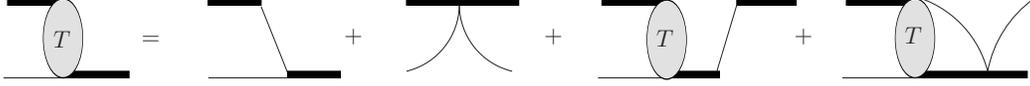}
\caption{The diagrammatic representation of the integral equation for the 1+2 scattering amplitude is shown.}
\label{fig:3btmat}
\end{center}
\end{figure}
\begin{align}
it(\vec{k},\vec{p},\varepsilon) = & -2g^2iS(-k^2/4m - B_2 + \varepsilon,\vec{p}+\vec{k}) + ih\notag \\
+&    \int_{-\infty}^\infty{\frac{dq}{2\pi}\frac{dq_0}{2\pi}}iS(k^2/2m - \varepsilon - q_0, -\vec{q})\notag \\
&  \left[(-2g^2)iS(-k^2/4m - B_2 + 2\varepsilon + q_0, \vec{p} + \vec{q})  + ih\right]\notag \\
& i\Delta(k^2/4m - B_2 + \varepsilon + q_0, \vec{q}) it(\vec{k},\vec{q},\varepsilon+q_0)
\end{align}
The $\int{\frac{dq_0}{2\pi}}$ integral is readily evaluated by contour integration since the two nucleon
propagators have poles in opposite half-planes.  The result is:
\begin{align}
it(\vec{k},\vec{p}) = & i2mg^2 \left\{\frac{1}{k^2 + p^2 - mE + \vec{p}\cdot\vec{k} - i\epsilon} + 
\frac{h}{2mg^2}\right\} \notag \\
-\:&  i 2mg^2  \int_{-\infty}^\infty{\frac{dq}{2\pi}}\left[ \frac{1}{q^2 + p^2 - mE + \vec{p}\cdot\vec{q}-i\epsilon} 
+ \frac{h}{2mg^2}\right]\notag \\
&  \frac{t(\vec{k},\vec{q})}{\Delta + \frac{mg^2}{2}[\frac{3q^2}{4} - mE - i\epsilon]^{-1/2}}  
\end{align}
With the chosen kinematics, the total energy is $E=3k^2/4m-B_2$.  We've set $\varepsilon=\frac{k^2-p^2}{2m}$,
and as in reference~\cite{BHvK}, defined $t(\vec{k},\vec{p})=t(\vec{k},\vec{p},\frac{k^2-p^2}{2m})$.  Now, averaging over the 
$p\cdot k$ brings us to:
\begin{align}
t(\vec{k},\vec{p}) = & 2mg^2\frac{k^2+p^2-mE}{(k^2 + p^2 - mE)^2 - p^2k^2} + h\notag \\
+&\:  4 \int_{-\infty}^\infty{\frac{dq}{2\pi}}\left[\frac{q^2+p^2-mE}{(q^2 + p^2 - mE)^2 - p^2q^2} + \frac{h}{2mg^2}\right] \notag \\
&  \frac{t(\vec{k},\vec{q})}{a_2 - [3/4(q^2-k^2)+mB_2-i\epsilon]^{-1/2}}  
\end{align}
The on-shell amplitude must include the wave-function normalization for the two-body bound state, which in 
field theory is conventionally written:
\begin{equation}
T_k=\sqrt{Z}t(k,k)\sqrt{Z}
\end{equation}
with
\begin{equation}
Z^{-1}=i\diff{}{p_0}(i\Delta(p))^{-1}\Big|_{p_0=-B_2} = \frac{mg^2}{2}\frac{-1}{2}\frac{-m}{(mB_2)^{-3/2}}=\frac{m^2g^2|a_2|^3}{4}
\end{equation}

It is desirable to bring this equation into the form of the standing-wave Lippman-Schwinger equation.
To this end, we define the function $a(k,p)$:
\begin{equation}
\frac{a(k,p)}{p^2-k^2}=\frac{t(k,p)/2mg^2}{a_2-(3p^2/4-mE)^{-1/2}}
\end{equation}
Since the integral is even in $q$ (indeed, it is only a function of $q^2$), the limits may be taken from 
zero to $\Lambda$  provided that we multiply by an overall factor of $2$.  This of course introduces
a sharp cutoff $\Lambda$.  The integral equation is now written in terms 
of a principal value prescription as:
\begin{equation}
\label{3b_ls}
a(k,p)=M(k,p;E) - \frac{4}{\pi} \mathcal{P}~\int_0^\Lambda{dq \: M(q,p;E)\frac{a(k,q)}{q^2-k^2}}
\end{equation}
where the kernel $M(q,p;E)$ is defined:
\begin{align}
M(q,p;E)=& \left[\frac{mE-q^2-p^2}{(mE-q^2-p^2)^2-q^2p^2} - \frac{h}{2mg^2}\right] \notag \\
&\left(\frac{4}{3a_2^2}\sqrt{-mE+3p^2/4}\left(1+a_2\sqrt{-mE+3p^2/4}\right)\right).
\end{align}
It is now clear that $a(k,p)$ satisfies the same integral equation as $K(k,p;E)$.


\bibliography{1D_del}

\end{document}